\documentclass[12pt,a4paper,titlepage]{article}
\usepackage{amsmath,color}
\usepackage{mathrsfs}

\def\slash#1{\not\!\!#1}
\renewcommand{\eqref}[1]{(\ref{#1})}

\usepackage{amsmath,amssymb}
\usepackage{graphicx}
\usepackage{here}
\definecolor{hyprf}{cmyk}{1,0.5,0,0}

\def\un#1{\relax\ifmmode\@@underline#1\else
        $\@@underline{\hbox{#1}}$\relax\fi}

\let\du=\du                     % dot-under
\def\m{\mu}
\def\n{\nu}

\def\F{\Phi}

\def\bo{{\raise-.3ex\hbox{\large$\Box$}}}               % D'Alembertian
\def\pa{\partial}                                       % curly d
\def\TH{{\raise.2ex\hbox{$\displaystyle \bigodot$}\mskip-4.7mu \llap H \;}}
\def\face{{\raise.2ex\hbox{$\displaystyle \bigodot$}\mskip-2.2mu \llap {$\ddot
        \smile$}}}                                      % happy face
\def\slash#1{\rlap{\hbox{$\mskip 1 mu /$}}#1}      % good slash for lower case
\def\Bar#1{\overline{#1}}                       % big bar
\def\leftrightarrowfill{$\mathsurround=0pt \mathord\leftarrow \mkern-6mu
        \cleaders\hbox{$\mkern-2mu \mathord- \mkern-2mu$}\hfill
        \mkern-6mu \mathord\rightarrow$}
\def\dvec#1{\vbox{\ialign{##\crcr
        \leftrightarrowfill\crcr\noalign{\kern-1pt\nointerlineskip}
        $\hfil\displaystyle{#1}\hfil$\crcr}}}           % <--> accent
\def\frac#1#2{{\textstyle{#1\over\vphantom2\smash{\raise.20ex
        \hbox{$\scriptstyle{#2}$}}}}}                   % fraction
\def\sfrac#1#2{{\vphantom1\smash{\lower.5ex\hbox{\small$#1$}}\over
        \vphantom1\smash{\raise.4ex\hbox{\small$#2$}}}} % alternate fraction
\def\bfrac#1#2{{\vphantom1\smash{\lower.5ex\hbox{$#1$}}\over
        \vphantom1\smash{\raise.3ex\hbox{$#2$}}}}       % "
\def\afrac#1#2{{\vphantom1\smash{\lower.5ex\hbox{$#1$}}\over#2}}    % "
\def\[{\lfloor{\hskip 0.35pt}\!\!\!\lceil}
\def\]{\rfloor{\hskip 0.35pt}\!\!\!\rceil}

\def\du#1#2{_{#1}{}^{#2}}

\def\fracm#1#2{\hbox{\large{${\frac{{#1}}{{#2}}}$}}}

\def\un{\underline}
\def\fracmm#1#2{{{#1}\over{#2}}}

\def\low#1{{\raise -3pt\hbox{${\hskip 0.75pt}\!_{#1}$}}}

\newskip\humongous \humongous=0pt plus 1000pt minus 1000pt

\newif\ifdtup

\def\({\left(}
\def\){\right)}

\def\beq{\begin{equation}}
\def\eeq{\end{equation}}
\def\bea{\begin{eqnarray}}
\def\eea{\end{eqnarray}}

\newcommand{\be}{\begin{equation}}
\newcommand{\ee}{\end{equation}}
\newcommand{\nbe}{\begin{equation*}}
\newcommand{\nee}{\end{equation*}}

\newcommand{\lb}{\label}

\begin{document}

\thispagestyle{empty}

{\hbox to\hsize{
\vbox{\noindent March 2020 \hfill IPMU20-0007}}
\noindent  revised version \hfill WU-HEP-20-01}

\noindent
\vskip2.0cm
\begin{center}

{\Large\bf Minimal Starobinsky supergravity coupled to dilaton-axion superfield}

\vglue.3in

Yermek Aldabergenov~${}^{a,b}$, Shuntaro Aoki~${}^{c}$, and Sergei V. Ketov~${}^{d,e,f}$
\vglue.3in

${}^a$~Department of Physics, Faculty of Science, Chulalongkorn University,\\
Thanon Phayathai, Pathumwan, Bangkok 10330, Thailand\\
${}^b$~Institute for Experimental and Theoretical Physics, Al-Farabi Kazakh National University,
71 Al-Farabi Avenue, Almaty 050040, Kazakhstan\\
${}^c$~Department of Physics, Waseda University, Tokyo 169-8555, Japan\\
${}^d$~Department of Physics, Tokyo Metropolitan University, \\
1-1 Minami-ohsawa, Hachioji-shi, Tokyo 192-0397, Japan \\
${}^e$~Research School of High-Energy Physics, Tomsk Polytechnic University,\\
2a Lenin Avenue, Tomsk 634028, Russian Federation \\
${}^f$~Kavli Institute for the Physics and Mathematics of the Universe (WPI),
\\The University of Tokyo Institutes for Advanced Study, Kashiwa 277-8583, Japan \\
\vglue.1in

 yermek.a@chula.ac.th, shun-soccer@akane.waseda.jp, ketov@tmu.ac.jp
\end{center}

\vglue.3in

\begin{center}
{\Large\bf Abstract}
\end{center}
\vglue.1in
\noindent The minimal Starobinsky supergravity with inflaton (scalaron) and goldstino in a massive vector supermultiplet is coupled to the dilaton-axion chiral superfield with the no-scale K\"ahler potential and a superpotential. The Kachru-Kallosh-Linde-Trivedi (KKLT)-type superpotential with a constant term is used to stabilize dilaton and axion during inflation, but it is shown to lead to an instability. The instability is cured by adding the alternative Fayet-Iliopoulos (FI) term that does not lead to the gauged $R$-symmetry.  Other stabilization mechanisms, based on the Wess-Zumino (WZ)-type superpotential, are also studied in the presence of the FI term. A possible connection to the D3-brane models is briefly discussed too.

\newpage

\section{Introduction}

Cosmological inflation offers a simple solution to basic problems of standard cosmology and current cosmological observations of the Cosmic Microwave Background (CMB) radiation.  Viable field theory models of  inflation are usually nonrenormalizable after quantization, which raises a problem of their Ultra-Violet (UV) completion in quantum gravity. It is important because physical predictions of those models are known to be sensitive to quantum corrections. Amongst all viable inflationary models, Starobinsky's model  \cite{Starobinsky:1980te} attracted a lot of attention because it provides (so far) the best fit to the cosmological observations \cite{planck2018}. The Starobinsky inflationary model of modified $(R+R^2)$ gravity and its scalar-tensor gravity counterpart are 
nonrenormalizable indeed, with the UV-cutoff being Planck mass $M_{\rm P}$ \cite{Burgess:2009ea,Hertzberg:2010dc}. Assuming quantum gravity to be given by string theory, the UV-completion in string theory implies the necessity to extend viable inflationary models to $N=1$ supergravity in four spacetime dimensions as the first step. A supergravity extension of the Starobinsky inflationary model is not unique, being dependent upon the supergravity framework chosen, see e.g., Ref.~\cite{Ketov:2019toi} for a review.  

The {\it minimal} description of Starobinsky inflation in supergravity as a single-field inflationary model (with a single physical scalar called scalaron) is possible when the scalaron (inflaton) is assigned to a massive abelian vector multiplet \cite{Farakos:2013cqa,Ferrara:2013rsa,Aldabergenov:2016dcu,Aldabergenov:2017bjt} in terms of unconstrained superfields. A generic action of a vector multiplet $V$ is governed by a single (real) potential $J(V)$, while its bosonic part in Einstein frame reads ($M_{\rm P}=1$)~\footnote{We use the spacetime signature $(-,+,+,+)$.}
\begin{equation} \label{complag}
e^{-1}\mathcal{L}_{\rm bos.}=\frac{1}{2}R -\frac{1}{4}F_{mn}F^{mn}-
\frac{1}{4}J_{CC}\partial_mC\partial^mC-\frac{1}{4}J_{CC}B_mB^m-\frac{g^2}{8}J_C^2~,
\end{equation}
where $R$ is Ricci scalar, $C$ is the leading field component of $V$, $B_m$ is an abelian vector field with the abelian
field strength $F_{mn}=\pa_mB_n-\pa_mB_n$ and the gauge coupling constant $g$, and the subscripts ($C$) of $J$ denote the derivatives of $J$ with respect to $C$. The scalar potential in Eq.~(\ref{complag}) is obtained after elimination of the auxiliary field $D$ of the vector multiplet, so that is of the $D$-type. The famous Starobinsky inflationary scalar potential is obtained by choosing the $J$ potential as
\begin{equation} \label{starj}
 J(C)=  -3\left( C + \ln(- C)\right)\quad {\rm with} \quad 
 C =  -\exp\left( \sqrt{2/3} \varphi\right)
 \end{equation}
in terms of the canonical scalaron $\varphi$. Then the first term in $J(C)$ is responsible for the induced cosmological constant driving inflation, whereas the second term in $J(C)$ represents an exponentially small correction during slow roll of the scalaron for positive values of $\varphi$, which is responsible for a suppression of the tensor-to-scalar ratio in the Starobinsky inflation, see e.g., Ref.~\cite{symmetryR} for details. 

Since inflation is driven by positive energy, in supergravity it leads to a Spontaneous Supersymmetry Breaking (SSB). Therefore, a {\it goldstino} should be present during inflation.  In the minimal supergravity description of inflation, the goldstino is given by gaugino ("photino") that is the superpartner of scalaron. It is worth to recall  that the goldstino action is {\it universal}, being given by the Akulov-Volkov (AV) action \cite{Volkov:1973ix} up to a field redefinition \cite{Hatanaka:2003cr,Kuzenko:2009ym}.  

Since an abelian vector multiplet is always present in the worldvolume of the spacetime-filling D3-brane (or anti-D3-brane)
\cite{Leigh:1989jq,Rocek:1997hi,Ketov:2001dq}, the D3-brane effective action may provide the desired embedding of the Starobinsky inflationary model into string theory. This conjecture is strongly supported by the existence of the Bagger-Galperin (BG) action \cite{Bagger:1996wp} of an $N=1$ abelian vector mulitplet, which is the Dirac-Born-Infeld-type (DBI-type) extension of the standard $N=1$ Maxwell-type action, because the BG action has the second (non-linearly realized) supersymmetry needed for its D-brane interpretation. However, in order to prove the conjecture, one needs to (i) realize Starobinsky's inflation in the DBI framework, (ii) provide SSB of the first (linearly realized) supersymmetry, and (iii) restore the second (non-linearly realized) supersymmetry after coupling  the BG (or DBI) action to supergravity.

The first problem was already solved in Refs.~\cite{Abe:2015fha,Aldabergenov:2018qhs}. The viable SSB  {\it after} the 
Starobinsky inflation, which gives rise to an adjustable (or observable) cosmological constant, is also possible by the use of the
alternative Fayet-lliopoulos (FI) terms without the gauging of the $R$-symmetry \cite{Cribiori:2017laj,Kuzenko:2018jlz,Aldabergenov:2018nzd,Abe:2018plc}. However, an origin of those FI terms in string theory and a restoration of another supersymmetry are still unclear, see e.g., Refs.~\cite{Cribiori:2018dlc,Antoniadis:2019xwa} for recent developments.

In this paper we do not address those unsolved problems but check whether the minimal formulation of the Starobinsky inflation in supergravity is compatible with its coupling to the chiral dilaton-axion superfield $\Phi$, described by the 
{\it no-scale} K\"ahler potential $K$ and a superpotential $W$. The no-scale K\"ahler potential reads
\be \lb{noscale}
K(\Phi,\Bar{\Phi}) = - n\log \left (\Phi + \Bar{\Phi}\right)~,
\ee
where we have introduced the real parameter $n>0$. In the context of string theory, the no-scale K\"ahler potential arises in toroidal (and orbifold) compactifications of type II strings and in the large volume limit of Calabi-Yau compactifications of heterotic strings (specifically, with $n=1$ for dilaton-axion and with $n=3$ for a volume modulus).The superpotential of dilaton and  axion in string theory may only be generated non-perturbatively. It is common in the literature to assume its specific form either as a Wess-Zumino-type cubic polynomial or as an exponential of $\Phi$.

At first sight, adding those couplings is not a problem in supergravity. However, we find that it spoils the Starobinsky inflation because of an instability. This phenomenon was first observed in Ref.~\cite{Aldabergenov:2017hvp} in the context of the
so-called Polonyi-Starobinsky supergravity where a Polonyi chiral superfield with the {\it canonical} kinetic term and a {\it linear} superpotential were introduced for describing SSB and dark matter  after inflation \cite{Addazi:2017ulg,Abe:2018rnu,Ketov:2019mfc} towards combining our (early time) inflationary models with  late time cosmology. In the case of the no-scale K\"ahler potential, we find a different  situation because the dilaton-axion has to be trapped near a minimum of their scalar potential during the Starobinsky
inflation driven by the scalaron, i.e., the masses of both dilaton and axion have to be larger than the Hubble scale during inflation (it is known as the moduli stabilization in the literature \cite{Douglas:2006es}). It is the purpose of this paper to achieve the moduli stabilization of dilaton and axion with the K\"ahler potential (\ref{noscale}) by using a suitable superpotential and the alternative FI term in the minimal Starobinsky supergravity coupled to the dilaton-axion superfield.~\footnote{We call $\F$ to be the dilaton-axion superfield for simplicity, although it may also represent a moduli superfield, with a generic parameter $n>0$.}  Our setup and motivation are different from those of Ref.~\cite{Antoniadis:2019xwa} where inflaton is identified with dilaton. They are also different from those of Ref.~\cite{Aldabergenov:2019aag}, where the superpotential is chosen in Polonyi's form.

Though we did our calculations with the DBI kinetic terms for the vector multiplet, in this paper we only use the Maxwell-type kinetic terms for simplicity, because the DBI structure does not significantly affect the Starobinsky inflation and the moduli stabilization in question, according to our findings (see also 
Refs.~\cite{Abe:2015fha,Abe:2018plc,Abe:2018rnu} for more).

The paper is organized as follows. In Sec.~2 we recall the superconformal tensor calculus in $N=1$ supergravity and introduce our notation. In Sec.~3 we define our inflationary model of the minimal Starobinsky supergravity, whose vector (inflaton) multiplet is coupled to the dilaton-axion chiral superfield with the no-scale K\"ahler potential. The vacuum structure of the model with the Kachru-Kallosh-Linde-Trivedi (KKLT)-type superpotential towards stabilization of dilaton and axion is investigated in Sec.~4, where an instability of inflation is found. A cure to the instability is proposed in Sec.~5 by using the alternative FI term. Sec.~6 is our conclusion. In Appendix A we study a different stabilization mechanism by using the Wess-Zumino (WZ)-type superpotential. Another stabilization mechanism,
 as a combination of the previous ones, is proposed in Appendix B. Spontaneous supersymmetry breaking after inflation is studied in Appendix C.

\section{Superconformal tensor calculus}

The conformal $N=1$ supergravity techniques are described in Refs.~\cite{Kaku:1978nz,Kaku:1978ea,Townsend:1979ki, Kugo:1982cu,Kugo:1983mv}. We follow the notation and conventions of Ref.~\cite{Freedman:2012zz}. In addition to the local symmetries of Poincar\'e supergravity, one also has the gauge invariance under dilatations, conformal boosts and
$S$-supersymmetry, as well as under $U(1)_A$ rotations. The gauge fields of dilatations and   $U(1)_A$ rotations are denoted by $b_{\mu}$ and $A_{\mu}$, respectively. A multiplet of conformal supergravity has charges with respect to dilatations and $U(1)_A$ rotations, called Weyl and chiral weights, respectively, which are denoted by pairs 
$({\rm{Weyl\ weight, chiral\ weight}})$ in what follows. 

A chiral multiplet has the field components 
\begin{align}
S=\{S,P_L\chi, F\}, \label{chiral}
\end{align}
where $S$ and $F$ are complex scalars, and $P_L\chi$ is a left-handed Weyl fermion ($P_L$ is the chiral projection operator).  In this paper  we use the two types of chiral multiplets: the conformal compensator $S_0$ and the matter multiplets $S^i$, where the index $i=1, 2 ,3 ,\ldots$, counts the matter multiplets.  The $S_0$ has the weights $(1,1)$ and is used to fix some of the superconformal symmetries. The matter multiplets $S^i$ have the weights $(0,0)$. The anti-chiral multiplets are denoted by $\bar{S}_0$ and $\bar{S}^{\bar{i}}$. 

As regards a general (real) multiplet, it has  the field content
\begin{align}
{\cal V}=\{ \mathcal{C}, \mathcal{Z}, \mathcal{H}, \mathcal{K}, \mathcal{B}_{a}, \Lambda , \mathcal{D}\} , \label{general}
\end{align}
where $ \mathcal{Z}$ and $\Lambda $ are fermions, $\mathcal{B}_a$ is a (real) vector, and others are (real) scalars, respectively.

The (gauge) field strength multiplet $W$ has the weights $(3/2,3/2)$ and the following field components: 
\begin{align}
\bar{\eta } W=\left\{ \bar{\eta }P_L\lambda ,\frac{1}{\sqrt{2}}\left(  -\frac{1}{2}\gamma _{ab}\hat{F}^{ab}+iD\right) P_L\eta  ,\bar{\eta  } P_L\slash{D}\lambda \right\} ,
\end{align}
where $\eta $ is the dummy spinor, $\hat{F}_{ab}=\partial_aB_b-\partial_bB_a+\bar{\psi}_{[a}\gamma_{b]}\lambda \equiv F_{ab}+\bar{\psi}_{[a}\gamma_{b]}\lambda$ is the  superconformally covariant field strength, the $\psi_a$ is gravitino, the $\lambda $ and $D$ are Majorana fermion and the real auxiliary scalar, respectively. The related expressions of the multiplets $W^2$ and $W^2\bar{W}^2$, which are embedded into the chiral multiplet~$\eqref{chiral}$ and the general multiplet~$\eqref{general}$, respectively, are 
\begin{align}
&W^2=\left\{ \cdots , \cdots, \cdots+\frac{1}{2} (FF-F\tilde{F})-D^2\right\} ,\\
&W^2\bar{W}^2=\left\{ \cdots , \cdots, \cdots,\cdots , \cdots, \cdots, \cdots+\frac{1}{2}| (FF-F\tilde{F})-2D^2|^2\right\} ,
\end{align}
where we have omitted the fermionic terms (denoted by dots) for simplicity. 
In addition, we use the book-keeping notation  $FF=F_{ab}F^{ab}$ and $\tilde{F}^{ab}\equiv -\frac{i}{2}\varepsilon ^{abcd}F_{cd}$ throughout the paper.

We also need another chiral multiplet
\begin{align}
\nonumber &\Sigma \left( \bar{W}^2/|S_0|^4\right) = \\
& \left\{ -\fracmm{ (\frac{1}{2}FF+\frac{1}{2}F\tilde{F}-D^2)}{|S_0|^4} +\cdots , \cdots, \fracmm{F_0}{|S_0|^4S_0}(FF+F\tilde{F}-2D^2)+\cdots\right\} ,  \label{D2W2}
\end{align}
where $\Sigma$ is the chiral projection operator \cite{Kugo:1982cu, Kugo:1983mv}. The argument of $\Sigma$ requires specific Weyl and chiral weights: in order to make sense to $\Sigma{\cal V}$, the ${\cal V}$ must satisfy $w-n=2$, where $(w,n)$ are the Weyl and chiral weights of ${\cal V}$. We adjust the correct weights of the argument by inserting the factor $|S_0|^4$. Equation~$\eqref{D2W2}$ is the conformal  supergravity counterpart of the superfield $\bar{D}^2\bar{W}^2$. 

The covariant derivative of $W$ is given by \cite{Kugo:1983mv}
\begin{align}
\mathcal{D}W=\{ -2D, \cdots , \cdots, \cdots,\cdots , \cdots,\cdots \} 
\end{align}
of the weights $(2,0)$. Here, the dots in the higher components also include some bosonic terms, but we do not write down them here for simplicity (see Ref.~\cite{Cribiori:2017laj} for their explicit expressions).

A massive vector multiplet $V$ has the field components 
\begin{align}
V=\{C,Z,H,K,B_a, \lambda ,D\}~,
\end{align}
while all of them are either real (bosonic) or Majorana (fermionic). The weights of $V$ are $(0,0)$.

The bosonic parts of the F-term invariant action formulas are 
\begin{align}
[S]_F=\frac{1}{2} \int d ^{4}x\sqrt{-g} \left( F+\bar{F}\right), \label{Fformula}
\end{align}
while they can be applied only when $S$ has the weights $(3,3)$. The bosonic part of the D-term formula for a real multiplet $\phi $ of the weights $(2,0)$ is 
\begin{align}
 [\phi ]_D= \int d^{4}x\sqrt{-g}\left( D_{\phi}-\frac{1}{3}C_{\phi}R(\omega)\right) ,\label{Dformula}
\end{align}
where $R(\omega)$ is the superconformal Ricci scalar in terms of spacetime metric and $b_{\mu}$~\cite{Freedman:2012zz}. The  $C_{\phi}$ and $D_{\phi}$ are the first and the last components of $\phi$, respectively.

We set the (reduced) Planck constant $M_{\rm P}$ and the abelian gauge coupling constant $g$ to unity for simplicity in our calculations, unless it is stated otherwise. Both of them can be restored by dimensional considerations and rescaling of the vector multiplet fields, respectively.

%%%%%%%%%%%%%%%%%%%%%%%%%%%%%%%%%%%%%%%%%%%%%%%%%%%%%%%%%%%%%%%%%%%%%%%%%%%%%%%%%%%%%%%%%

\section{The model}

Let us consider the supergravity model of a massive vector multiplet $V$ coupled to a dilaton-axion chiral multiplet $\Phi$, whose  action is given by
\begin{align}
S=-\frac{3}{2}\biggl[ |S_0|^2e^{-\mathcal{J}/3}\biggr] _D+2[S_0^3\mathcal{W}]_F-[\Phi W^2]_F ~~, \label{total action}
\end{align}
where $\mathcal{J}$ is a real function of $C$ and $(\Phi+\bar{\Phi})$. We take  $\mathcal{J}$ as a sum of the
Starobinsky potential (\ref{starj}) and the no-scale K\"ahler potential (\ref{noscale}),
\begin{align} \lb{master}
\mathcal{J}=-3\log \left(  -Ce^{C}\right)-n \log (\Phi+\bar{\Phi}),
\end{align}
where $n$ is a positive integer. The first term in Eq.~(\ref{master})  is supposed to be responsible for the Starobinsky-type inflation, and the second one describes the interactions of dilaton and axion. The $\mathcal{W}$ is a holomorphic superpotential depending on $\Phi$ only.

Our action is invariant under a constant shift 
\begin{align}
\Phi \rightarrow \Phi+ic \label{shift}
\end{align}
with a real constant $c$, except of the superpotential term. 

After imposing the superconformal gauge fixing and integrating out the auxiliary fields, the bosonic part of the action
$\eqref{total action}$ is given by 
\begin{align}
\nonumber  \mathcal{L}=&\frac{1}{2}R-\frac{1}{4}\mathcal{J}_{CC}(\partial_aC)^2 -\frac{1}{4}\mathcal{J}_{CC}B_a^2-\mathcal{J}_{\Phi\bar{\Phi}}\partial_a\Phi\partial^a\bar{\Phi}-V\\
&-\frac{1}{4}(\Phi+\bar{\Phi})FF+\frac{1}{4}(\Phi-\bar{\Phi})F\tilde{F}.\label{L total}
\end{align}
The subscripts of $\mathcal{J}$ denote the derivatives of $\mathcal{J}$ with respect to the scalar fields $C$, 
$\Phi$ and $\bar{\Phi}$, respectively. The $V$ is the scalar potential, whose explicit form reads  
\begin{align}
&V=V_F+V_D~~,\\
&V_F=e^{\mathcal{J}}\biggl[(\mathcal{J}_{\Phi\bar{\Phi}})^{-1}|\mathcal{W}_{\Phi}+\mathcal{J}_{\Phi}\mathcal{W}|^2+\left(\fracmm{\mathcal{J}_C^2}{\mathcal{J}_{CC}}-3\right)|\mathcal{W}|^2\biggr]~~,\\
&V_D=\fracmm{\mathcal{J}_C^2}{8(\Phi+\bar{\Phi})}~~,
\end{align}
in agreement with Refs.~\cite{Aldabergenov:2016dcu,Aldabergenov:2017bjt,Aldabergenov:2019aag}. When $\mathcal{W}=0$, the Lagrangian (\ref{L total}) reduces to the one of Ref.~\cite{Ketov:2003gr}  considered in the context of global supersymmetry. In the absence of $\Phi$, the equations above reduce to Eqs.~(\ref{complag}) and (\ref{noscale}). 

The Planck mass $M_{\rm{P}}$ can be recovered as follows: 
\begin{align}
\fracmm{C}{M_{\rm{P}}}=-e^{\sqrt{\frac{2}{3}}\frac{\varphi}{M_{\rm{P}}}}~~, \quad \fracmm{\Phi}{M_{\rm{P}}}=e^{-\sqrt{\frac{2}{n}}\frac{\phi}{M_{\rm{P}}}}+i \sqrt{\fracmm{2}{n}}\fracmm{a}{M_{\rm{P}}}~~. \label{canonical}
\end{align}
The fields $\varphi,\phi$ and $a$ can be identified as the canonical inflaton/scalaron, dilaton, and axion, respectively. After rewriting the Lagrangian in terms of those fields, 
we obtain
\begin{align}
\mathcal{L}=&-\frac{1}{2}(\partial_a\varphi)^2 -\frac{1}{2}(\partial_a\phi)^2-\frac{1}{2}e^{2\sqrt{\frac{2}{n}}\frac{\phi}{M_{\rm{P}}}}(\partial_aa)^2-V_F-V_D,
\end{align}
where (after a restoration of the gauge coupling constant $g$ also) we have
\begin{align}
\nonumber V_F=&e^{\mathcal{J}/M_{\rm{P}}^2}\biggl[\fracmm{e^{-2\sqrt{\frac{2}{n}}\frac{\phi}{M_{\rm{P}}}}}{n}|\mathcal{W}_{\Phi}|^2-e^{-\sqrt{\frac{2}{n}}\frac{\phi}{M_{\rm{P}}}}\left(\mathcal{W}_{\Phi}\fracmm{\bar{\mathcal{W}}}{M_{\rm{P}}}+\bar{\mathcal{W}}_{\bar{\Phi}}\fracmm{\mathcal{W}}{M_{\rm{P}}}\right)\\
&+\left(n-6e^{\sqrt{\frac{2}{3}}\frac{\varphi}{M_{\rm{P}}}}+3e^{2\sqrt{\frac{2}{3}}\frac{\varphi}{M_{\rm{P}}}}\right)
\fracmm{|\mathcal{W}|^2}{M_{\rm{P}}^2}\biggr] ~~,\\
V_D=&  \fracmm{9g^2M_{\rm{P}}^4}{16}e^{\sqrt{\frac{2}{n}}\frac{\phi}{M_{\rm{P}}}}\left(1-e^{-\sqrt{\frac{2}{3}}\frac{\varphi}{M_{\rm{P}}}}\right)^2.
\end{align}

The Starobinsky inflation is supposed to be driven by the $D$-type term above. However, in the case under consideration the $D$-term has the dilaton-dependent factor. Therefore, a viable inflation is only possible after a stabilization of the dilaton, while keeping the $F$-term to be relatively small against the $D$-term, so that the $F$-term should not spoil the Starobinsky inflation either.

\section{The vacuum structure}\label{stabilize}

We choose the Kachru-Kallosh-Linde-Trivedi (KKLT)-type superpotential  \cite{Kachru:2003aw} as our Ansatz to safely stabilize dilaton and axion in our model (Sec.~3).

Let us first study the stationary conditions of all fields. In terms of the inflaton~$C=-e^{\sqrt{\frac{2}{3}}\varphi}$, the dilaton ${\rm{Re}}\,\Phi=\rho $, and the axion ${\rm{Im}}\,\Phi=\theta$,\footnote{Canonical normalizations of the dilaton and axion fields are  obtained after a field redefinition $\Phi=e^{-\sqrt{\frac{2}{n}}\phi}+i\sqrt{\frac{2}{n}}a$. We find convenient to use $\rho$ and $\theta$ in what follows.} the scalar potential is given by
\begin{align}
V_D=& \fracmm{9 g^2}{16 \rho } \left(1- e^{-\sqrt{\frac{2}{3}} \varphi }\right)^2,\label{VD}\\
 V_F=&e^J\fracmm{1}{(2\rho)^n}\biggl[\fracmm{(2\rho)^2}{n}|W_{\Phi}|^2-2\rho \left(W\bar{W}_{\bar{\Phi}}+\bar{W}W_{\Phi}\right)+(P+n)|W|^2\biggr],\label{VF}
\end{align}
where we have introduced the notation
\begin{align}
&J(\varphi)=3 e^{\sqrt{\frac{2}{3}} \varphi }-\sqrt{6} \varphi~,\label{def_J}\\
&P(\varphi)=3 e^{\sqrt{\frac{2}{3}} \varphi } \left(e^{\sqrt{\frac{2}{3}} \varphi }-2\right) \label{def_P}~~,
\end{align}
and have recovered the gauge coupling constant $g$ that determines the inflationary scale.

The first derivative of the scalar potential with respect to $\varphi$ reads
\begin{align}
 V_{\varphi}=& \fracmm{9 g^2}{8 \rho }\sqrt{\frac{2}{3}} \varphi\left(1- e^{-\sqrt{\frac{2}{3}} \varphi }\right)+J_{\varphi}V_F+e^J\fracmm{1}{(2\rho)^n}P_{\varphi}|W|^2~~,\label{Vvarphi} 
\end{align}
where the subscripts denotes the derivatives with respect to a given field. The $\varphi=0$ is a solution of $V_{\varphi}=0$, since $J_{\varphi}$ and $P_{\varphi}$ vanish at $\varphi=0$. 
%The other stationary conditions for $\rho$ and $\theta$ are model dependent, thus we will investigate the two cases with the different superpotentials in the following.

Let us assume that the superpotential takes the following form:
\begin{align}
\mathcal{W}=W_0+A e^{-B \Phi} \label{super_KKLT}
\end{align}
that is inspired by the KKLT-type superpotential \cite{Kachru:2003aw} with constant parameters $W_0$, $A$ and $B$.~\footnote{Another type of the superpotential is considered in Appendix~\ref{WZ type}.} The non-vanishing constant $W_0$ is essential in our investigation. We assume that $W_0$ is negative and $A, B$ are both positive. In this case, the $F$-term scalar potential is explicitly given by
\begin{align}
\nonumber V_F=&e^J\fracmm{1}{(2\rho)^n}\biggl[\fracmm{4A^2B^2\rho^2}{n}e^{-2B\rho}+4 A B \rho  e^{-2 B \rho } \left(A+W_0 e^{B \rho } \cos (B\theta)\right)\\
&+(P+n)\left(W_0^2+2 A W_0 e^{-B \rho } \cos (B\theta)+A^2 e^{-2 B \rho }\right)\biggr].
\end{align}
We find that $\theta=2m\pi,~m\in \mathbb{Z}~,$ minimizes the potential for $n\geq 3$ since $P+n\geq 0$ holds in that case. In the following, we focus on the point at $\theta=0$. Taking into account the condition~$\varphi=0$ for $V_{\varphi}=0$, the condition $V_{\rho}=0$ is reduced to either of  
\begin{align}
&A (2 B \rho +n)+n W_0 e^{B \rho }=0, \label{Vrho1}\\ 
&A \left(n (4 B \rho -3)+4 B \rho  (B \rho -1)+n^2\right)+(n-3) n W_0 e^{B \rho }=0.\label{Vrho2}
\end{align}
These conditions should be regarded as the equations that determine the vacuum expectation value of $\rho~(=\rho_0)$.

In what follows, we consider the no-scale case with $n=3$ for definiteness. Then Eq.~$\eqref{Vrho1}$ becomes
\begin{align}
W_0=-Ae^{-B\rho_0}\left(1+\frac{2}{3}B\rho_0\right)
\end{align}
that is exactly same as the KKLT vacuum.~\footnote{Equation $\eqref{Vrho2}$ yields the solutions 
$\rho=0,-\frac{2}{B}$~.} The relevant masses at the stationary point are explicitly given by
\begin{align}
\nonumber &m^2_{\varphi}=\fracmm{27 g^2-4 A^2 B^2 e^{3-2 B \rho_0}}{36\rho_0},\\
&m^2_{\rho}=\fracmm{A^2 B^2 e^{3-2 B \rho_0} (B \rho_0+2) (2 B \rho_0+1)}{6 \rho_0^3},\label{mass_KKLT} \\
\nonumber&m^2_{\theta}=\fracmm{A^2 B^3 e^{3-2 B \rho_0} (2 B \rho_0+3)}{6 \rho_0^2},
\end{align}
and they are all {\it positive}.~\footnote{We assume that the inflationary scale $\sim g$ is
larger than that of $V_F$.} The minimum is of the Anti-de-Sitter (AdS)-type because the cosmological constant is given by 
\begin{align} \label{adsv}
V_0=-\fracmm{A^2 B^2 e^{3-2 B \rho_0}}{6 \rho_0} < 0~, 
\end{align}
while supersymmetry is restored at the minimum.

Though both dilaton and axion are stabilized by using the KKLT-type superpotential, as was demonstrated above, there is still a problem. In the Starobinsky-type inflationary scenario, it is necessary to require $V_D\gg |V_F|$. But the double exponential in the $e^J$-factor and the exponentials in the $P$ function, defined by Eqs.~(\ref{def_J}) and (\ref{def_P}) in terms of the {\it canonical} inflaton $\varphi$, destroy the flatness of  the scalar potential and thus greatly reduce the e-foldings number of inflation. Therefore, we need the hierarchy of the two parameters, namely, $g\gg A$. However, as can be seen from Eq.~$\eqref{mass_KKLT}$, it gives rise to the extremely small dilaton and axion masses and, therefore,  the KKLT stabilization Ansatz alone does not work here.~\footnote{The range of the scalaron field $\varphi/M_{\rm P}$ is trans-Planckian of the order ${\cal O}(1)$ during the (large field) Starobinsky inflation \cite{Aldabergenov:2018qhs}.}

%\subsection{Introduction of FI term}
%To resolve the problem, we introduce the following FI term~\cite{Cribiori:2017laj},
%\begin{align}
%S_{FI}=-\frac{3}{2}\biggl[ |S_0|^2e^{-\mathcal{J}/3}\xi \frac{W^2\bar{W}^2}{(\mathcal{D}W)^2(\bar{\mathcal{D}}\bar{W})^2}\mathcal{D}W\biggr] _D , \label{S_FI}
%\end{align}
%where $\xi$ is a real function and we assume that it depends on $C$ and a combination $\Phi+\bar{\Phi}$, not to break Eq.~$\eqref{shift}$. The FI term does not require a gauged R-symmetry and is applicable for the two types of superpotential discussed in the last section. Here we discuss the WZ type superpotential with $n=4$. 

%As for the scalar sector, only the modification appears in the D-term potential as 
%\begin{align}
%V_D=& \frac{9 g^2}{16 \rho } \left(1- e^{-\sqrt{\frac{2}{3}} \varphi }-\frac{\mathcal{\xi}}{9g}\right)^2.
%\end{align}
%When FI term is introduced, we cannot obtain an analytical expression of the stationary point. Thus, we specify the deviations as $\varphi=0+\delta \varphi$, $\rho=\rho_0+\delta \rho$, and $\theta=\rho_0+\delta \theta$, and expand the scalar potential as

%%%%%%%%%%%%%%%%%%%%%%%%%%%%%%%%%%%%%%%%%%%%%%%%%%%%%%%%%%%%%%%%%%%%%%%%%%%%%%%%%%%%%%%%%%%%%%%%%%%%%%%%%%%%%%%%%%%%%%%%%%%%%%%%%%%%%%%%%%%%%%%%%%%%%%%%%%%%%%%%

\section{The field-dependent FI term}\label{newFI}

As long as the total $J$-potential $\eqref{starj}$ is unchanged, it gives rise to the Starobinsky $D$-type inflationary potential, as desired. However, as we found in the previous Section, the coupling of the vector (inflaton) superfield to the chiral (dilaton-axion) superfield converts a single-field inflation into a multi-field inflation, while it leads to the instability resulting in a significant reduction of the duration of inflation, measured by the e-foldings number $N_e$, and, hence, to an unacceptable change in the predicted CMB power spectrum measured by the scalar index $n_s$ and the tensor-to-scalar ratio $r$, both depending upon $N_e$. Our idea to solve this problem is to change the origin of the first term in the $J$-potential $\eqref{starj}$
 and thus avoid the instability via the induced change in the $F$-type scalar potential.

Let us introduce the following {\it alternative} field-dependent FI term ({\it cf.} Refs.~\cite{Cribiori:2017laj,Kuzenko:2018jlz,Aldabergenov:2018nzd}):~\footnote{The standard FI term \cite{Fayet:1974jb} in the context of supergravity does not work because it implies the gauged $R$-symmetry and, hence, the charged gravitino field that severely restricts possible couplings.}
\begin{align}
S_{FI}=-\fracm{3}{2}\biggl[ |S_0|^2e^{-\mathcal{J}/3}\xi \fracmm{W^2\bar{W}^2}{(\mathcal{D}W)^2(\bar{\mathcal{D}}\bar{W})^2}\mathcal{D}W\biggr] _D~~, \label{S_FI}
\end{align}
where $\xi$ is a real function that, in general, depends on $C$ and a combination $(\Phi+\bar{\Phi})$, in order to preserve the shift symmetry in Eq.~$\eqref{shift}$. This FI term does not require the gauged R-symmetry, and therefore, is applicable together with our KKLT-type superpotential.~\footnote{The gauged 
$R$-symmetry does not allow a constant term in the superpotential.}  It appears that a constant $\xi$ does not help because it merely shifts the vacuum and does not contribute to $e^{J}$ and $P$. As was noticed in Ref.~\cite{Aldabergenov:2017hvp}, the dangerous terms in the scalar potential can be removed when $\xi$ and $J$ satisfy a specific relation, by extending $\xi$ to be field dependent. Here we apply the same idea to the case under consideration, where the dilaton-axion multiplet is coupled to the massive vector multiplet.  

Let us choose $J$ and $\xi$ so they satisfy the relation
\begin{align}
J_C+\fracmm{\xi(C)}{3g}=-3\left(1+\fracmm{1}{C}\right)~,\label{deom}
\end{align}
where we have set 
\begin{align}
\xi(C)=3g\xi_0e^{kJ} \left(1+\fracmm{1}{C}\right), \quad \xi_0<0, \quad k>0~. \label{def_xi}
\end{align}
The case $\xi(C)\propto e^{kJ} $ was studied in Ref.~\cite{Aldabergenov:2017hvp}. In Eq.~(\ref{def_xi})
 we added the factor $\left(1+\frac{1}{C}\right)$ to ensure $\xi(-1)=0$. This factor does not change the results of Ref.~\cite{Aldabergenov:2017hvp} since it is reduced to $1$ for $C\rightarrow -\infty$. However, due to a change of the $J$-function in Eq.~$\eqref{deom}$, the canonical scalaron field in 
 Eq.~$\eqref{canonical}$ has to be modified. We find convenient to use the non-canonical inflaton field $C$ in what follows. 
 
For large negative $C$, Eq.~$\eqref{deom}$ can be approximately solved as 
\begin{align}
J\sim -\frac{1}{k}\log \frac{1}{3}\left(e^{3k(C-C_0)}-\xi_0\right), \label{sol_deom}
\end{align}
with the integration constant $C_0$. Thus the function $J$ becomes constant as 
\begin{align}
J_{\infty}\equiv -\fracmm{1}{k}\log \fracmm{-\xi_0}{3},\label{Jinfty}
\end{align}
for $C\rightarrow -\infty$. Then the exponential factor $e^J$ in the $F$-type part $V_F$ of the scalar potential $\eqref{VF}$ also becomes constant 
during inflation, while the term $P(C)=J_C^2/J_{CC}-3$  in the $V_F$ becomes constant too,
\begin{align}
P_{\infty}=\fracmm{3-2\xi_0e^{kJ_{\infty}}}{k\xi_0e^{kJ_{\infty}}}~~.
\end{align}

To summarize, we obtain the following full scalar potential during inflation:
\begin{align}
\nonumber V=&\fracmm{9g^2}{16\rho}\left(1+\fracmm{1}{C}\right)^2\\
&+e^{J_{\infty}}\fracmm{1}{(2\rho)^n}\biggl[\fracmm{(2\rho)^2}{n}|W_{\Phi}|^2-2\rho \left(W\bar{W}_{\bar{\Phi}}+\bar{W}W_{\Phi}\right)+(P_{\infty}+n)|W|^2\biggr]~,
\end{align}
whose $F$-term in the second line does not spoil the Starobinsky inflation described by the $D$-term in the first line because $e^{J_{\infty}}$ and $P_{\infty}$ are constants during the inflation. The observational predictions for the cosmological tilts $n_s$ and $r$ with the e-folding number $N_e$ between 50 and 60 
 in this inflationary model are the same as in the Starobinsky case, see Ref.~\cite{Aldabergenov:2017hvp} for details.

As the inflation ends, the inflaton $C$ and the dilaton-axion $\Phi=\rho +i \theta$ take the vacuum expectation values which are determined by the vacuum conditions $V_C=V_{\rho}=V_{\theta}=0$. We find that $C=-1$ is still a solution to $V_C=0$ since $J_C|_{C=-1}=P_C|_{C=-1}=0$ in the parameterization of Eq.~$\eqref{def_xi}$. As regards $V_{\rho}=V_{\theta}=0$, the results of the previous Sec.~\ref{stabilize} apply since $V_D|_{C=-1}=0$. 

Moreover, we do {\it{not}\/} have to demand $V_D\gg |V_F|$ with the FI term because  the structure of $J$ and $\xi$ already solves the problem. Therefore, we can strongly stabilize $\Phi$ by choosing the superpotential parameters appropriately, i.e., with a sufficiently large $A$ in Eq.~$\eqref{super_KKLT}$.  

For completeness, we provide the masses of all scalar fields in the vacuum. The masses given in Eq.~$\eqref{mass_KKLT}$ get small modifications due to 
the change  of the $J$-function, see Eq.~$\eqref{deom}$. They are explicitly given by
\begin{align}
\nonumber &m^2_{C}=\fracmm{9g^2}{8\rho_0}-\fracmm{A^2 B^2 e^{J(-1)-2 B \rho_0}}{6 \rho_0}\left(1+\frac{1}{3}\xi_0e^{kJ(-1)}\right)~,\\
&m^2_{\rho}=\fracmm{A^2 B^2 e^{J(-1)-2 B \rho_0} (B \rho_0+2) (2 B \rho_0+1)}{6 \rho_0^3}~~,\label{mass_KKLT_2} \\
\nonumber&m^2_{\theta}=\fracmm{A^2 B^3 e^{J(-1)-2 B \rho_0} (2 B \rho_0+3)}{6 \rho_0^2}~~,
\end{align}
where $J(-1)$ is the value at $C=-1$. In the limit $\xi_0\rightarrow 0$, these masses are reduced to those in Eq.~$\eqref{mass_KKLT}$.

A comment is in order here. Consistency of the alternative FI term (\ref{S_FI}) requires the vacuum expectation value of the $D$-term to be nontrivial; in other words, supersymmetry must be spontaneously broken in the vacuum. Since the $D$ in our case is given by 
\begin{align}
D=\fracmm{g}{2\rho}\left(J_C+\fracmm{\xi(C)}{3g}\right)~,
\end{align}
the vacuum discussed above is not allowed. To be consistent, the right-hand-side of Eq.~$\eqref{deom}$ should be modified further as 
\begin{align}
J_C+\fracmm{\xi(C)}{3g}=-3\left(1+\fracmm{1}{C}\right)+\Delta~,\label{deom2}
\end{align}
where a non-vanishing (small) constant $\Delta$ has been introduced. As long as $|C \Delta|\ll 1$ during inflation, the proposed mechanism applies without changing the results above. The detailed analysis is given in Appendix~\ref{delta}.
%%%%%%%%%%%%%%%%%%%%%%%%%%%%%%%%%%%%%%%%%%%%%%%%%%%%%%%%%%%%%%%%%%%%%%%%%%%%%%%%%%%%%%%%%%%%%%%%%%%%%%%%%%%%%%%%%%%%%%%%%%%%%%%%%%%%%%%%%%%%%%%%%%%%%%%%%%%%%%%

\section{Conclusion}

In this paper we studied the phenomenological aspects of inflation in our new supergravity model. Our main results are summarized in the Abstract.

As we already mentioned in the Introduction, the DBI deformation of the vector multiplet kinetic terms in our supergravity model, which is essentially described by a locally supersymmetric extension of the BI action
\be \lb{bi}
-\frac{1}{4}\sqrt{-g}F_{\m\n}F^{\m\n} \to M_{\rm BI}^4 \left[ \sqrt{-\det(g_{\m\n})} - \sqrt{-\det\left(g_{\m\n}+
M_{\rm BI}^{-2}F_{\m\n}\right)} \,\right]~
\ee
with the dimensional deformation parameter $M_{\rm BI}^4$, is available and does not significantly change our results.
The DBI structure is, however, relevant for a possible embedding of our model into the effective action of a D3-brane.

The BI action is known to have the $U(1)$ electric-magnetic self-duality, while its minimal coupling to the {\it massless} dilaton and axion results in the 
$SL(2,{\bf R})$ self-duality \cite{Gaillard:1981rj} that also applies to the D3-brane effective action of the massless fields. The manifestly $N=1$ locally supersymmetric extension of the BI action, coupled to the massless dilaton-axion chiral superfield and preserving  the $SL(2,{\bf R})$ self-duality, can be found in Ref.~\cite{Kuzenko:2002vk}.~\footnote{See also the related construction in Ref.~\cite{Aldabergenov:2018vju} towards  the electromagnetic confinement.}  The 
self-duality properties are only valid in the case of the massless fields and in the absence of a superpotential.~\footnote{ Both dilaton and axion are massless in the perturbative string theory.} 

{\it Another} (non-linearly realized) supersymmetry is also required for a D3-brane. Our supergravity model has manifest $N=1$ supersymmetry but does not  have another supersymmetry by construction, though it may still be possible after a modification of our action or by using non-linear realizations where 
manifest supersymmetry is absent. Unlike the standard FI term, the alternative FI terms avoid the no-go theorems known in supergravity and string theory
\cite{Komargodski:2009pc} so that the search for an origin of the alternative FI term (\ref{S_FI}) in string theory deserves further investigation.

Finally, we mention a possible connection to extended supersymmetry and supergravity. Some alternative FI terms were recently found in $N=2$ supergravity \cite{Antoniadis:2019hbu}. The $N=2$ supersymmetric extensions of the BI theory both in superspace and via non-linear realizations also exist 
\cite{Ketov:1998ku,Ketov:1998sx,Kuzenko:2000uh,Bellucci:2000ft,Bellucci:2012nz}. The scalar ($\phi^i)$ kinetic terms of the $N$-extended vector multiplet  enter the generalized   BI action via the root
\be \lb{bis}
-M_{\rm BI}^4\sqrt{ -\det\left[ g_{\m\n}+ M_{\rm BI}^{-2}(F_{\m\n}+\partial_{\mu}\phi^i\partial_{\nu}\phi^i)\right]}~,
\ee
which is different from the $k$-inflation \cite{ArmendarizPicon:1999rj}  and Horndeski gravity theories \cite{Deffayet:2009wt}. 

\section*{Acknowledgements}

 The authors are grateful to Hiroyuki Abe for his collaboration on related work in Refs.~\cite{Abe:2018plc,Abe:2018rnu} and
 Sergei Kuzenko for correspondence. S.A. was supported in part by a Waseda University Grant for Special Research Projects (Project number: 2019E-059). Y.A.  was supported by the CUniverse research promotion project of Chulalongkorn University in Bangkok, Thailand, under the grant reference CUAASC, and by the Ministry of Education and Science of the Republic of Kazakhstan under the grant reference BR05236322. S.V.K. was supported by Tokyo Metropolitan University, the World Premier International Research Center Initiative (WPI), MEXT, Japan, and the Competitiveness Enhancement Program of Tomsk Polytechnic University in Russia.

\begin{appendix}

\section{The WZ-type superpotential} \label{WZ type}

Le us investigate another case of the Wess-Zumino (WZ)-type superpotential in order to stabilize dilaton and axion during Starobinsky inflation in our supergravity model. We did our calculations with a generic (cubic) WZ superpotential, but those results are cumbersome and not very illuminating. We restrict ourselves in this Appendix to the most relevant mass term
for simplicity, i.e.,
\begin{align}
W=m\Phi^2~,\label{WZSP}
\end{align}
where $m$ is a real constant. The $F$-term potential becomes
\begin{align}
 V_F=&e^J\fracmm{m^2}{(2\rho)^n}\left(\theta^2+\rho ^2\right)\biggl[  \left(\theta^2+\rho ^2\right) \left(n+P\right)+\fracm{16 \rho ^2}{n}-8  \rho ^2\biggr].
\end{align}
Under the condition~$\varphi=0$, we obtain 
\begin{align}
&n=3:\quad V_{\rho}= -\fracmm{e^3 m^2 \left(\rho ^2-\theta^2\right)}{3 \rho ^2}, \quad V_{\theta}= -\fracmm{2 e^3 \theta m^2}{3 \rho }~,\\
&n=4: \quad V_{\rho}= \fracmm{e^3 \theta^2 m^2 \left(\rho ^2-\theta^2\right)}{4 \rho ^5}, \quad V_{\theta}=  -\fracmm{e^3 \theta m^2 \left(\rho ^2-\theta^2\right)}{4 \rho ^4}~.
\end{align}

It is easy to verify that the case of $n=3$ has no solution. In the $n=4$ case, the equations are satisfied when 
$\rho=\theta\equiv \rho_0$. The masses are given by
\begin{align}
&m^2_{\varphi}= \fracmm{mm3 g^2}{4 \rho_0}+\fracmm{e^3 m^2}{2}~,\\
&m^2_{-}=\fracmm{e^3 m^2}{\rho_0^2}, \quad m^2_{+}=0~, \label{mass_WZ}
\end{align}
where $m^2_{\pm}$ are the masses of $\frac{1}{\sqrt{2}}(\rho \pm \theta)$. Hence, the vacuum is not stabilized in this case. 

In contrast to the KKLT case, supersymmetry is broken in the vacuum because
\begin{align}
F_{\Phi}\sim \fracmm{m}{\rho_0}~~. \label{F_Phi}
\end{align}
The vacuum is AdS, whose depth is given by
\begin{align}
V_0=-\frac{1}{4} e^3 m^2.    
\end{align}

The situation can be improved by adding {\it quartic} couplings inside the log of the K\"ahler potential in the $\mathcal{J}$ function. Let us modify $\mathcal{J}$ as
\begin{align}
\mathcal{J}=J(C)-n \log \left[ \Phi+\bar{\Phi}+\gamma_1(\Phi+\bar{\Phi}-2\rho_0)^4+\gamma_2(\Phi-\bar{\Phi}-2i\rho_0)^4\right]~,\label{J_gamma}
\end{align}
where $\gamma_{1,2}$ are the real parameters.  After these modifications, the stationary point is the same as in the model without the quartic modifications, 
i.e.,    
\begin{align}
\varphi=0,\quad \rho=\theta=\rho_0, \quad ({\rm{with}}\ n=4)~. \label{stationary_WZ}
\end{align}

The quartic couplings affect the mass terms that can be roughly evaluated as 
\begin{align}
\Delta m^2 \sim (\mathcal{J}_{\Phi \bar{\Phi}})^{-1}_{\Phi \bar{\Phi}}|F_{\Phi}|^2~.
\end{align}
Thus the messes get no corrections when the vacuum preserves supersymmetry, like in the KKLT case. In the 
WZ-type model, we find the contributions to the mass matrix as follows:
\begin{align}
\begin{pmatrix} 
M^2_{\varphi\varphi} & M^2_{\varphi\rho} & M^2_{\varphi\theta}\\ 
* & M^2_{\rho\rho}&M^2_{\rho\theta}\\ 
* & * &M^2_{\theta\theta}\\
\end{pmatrix} 
=
\left(
\begin{array}{ccc}
 \frac{3 g^2}{4 \rho_0}+\frac{e^3 m^2}{2} & 0 & 0 \\
 0 & \frac{e^3 m^2 \left(192 \gamma_1 \rho_0^3+1\right)}{2 \rho_0^2} & -\frac{e^3 m^2}{2 \rho_0^2} \\
 0 & -\frac{e^3 m^2}{2 \rho_0^2} & \frac{e^3 m^2 \left(192 \gamma_2 \rho_0^3+1\right)}{2 \rho_0^2} \\
\end{array}
\right).    \label{mass_Mat}
\end{align}
Therefore, we can stabilize $\rho$ and $\theta$ in the presence of the quartic couplings when the latter take values larger than $m$.  We can also decouple the masses of the dilaton and the axion from $V_F\sim m^2$, and impose the condition $V_D\sim g^2\gg V_F\sim m^2$ in this case. As regards inflation, the mechanism discussed in the main text (Sec.~5) can be applied here too.

\section{Hybrid solution}

Since the problem of suppressing the $V_F$ term comes from the exponential  in
\begin{equation}
    J=-3\log(-Ce^{C})~,\label{fullj}
\end{equation}
(in terms of the canonically normalized inflaton $\varphi$, we have  $C=-e^{\sqrt{\frac{2}{3}}\varphi}$), we can change the $J$ function to the first term only as
\begin{equation} \label{cutj}
    J=-3\log{(-C)}~,
\end{equation}
and generate the second term in (\ref{fullj}), leading to the constant vacuum energy driving inflation in the $D$-type Starobinsky potential and responsible for the instability due to the $F$-term, by the alternative FI term, schematically 
as $\sim (1/C+\xi)^2$. The parameter $\xi$ can be fixed as $\xi=g$ in order to keep the standard Starobinsky potential.

As regards the dilaton-axion coupled model, we have to demand the following conditions: (i) the D-term potential should not cross zero between the start of inflation and the vacuum (otherwise, the action will become singular due to the alternative FI term), and (ii) the masses of dilaton and axion must be higher than the inflationary (Hubble) scale during inflation.

Let us introduce the dilaton-axion pair with the following no-scale K\"ahler potential and the WZ superpotential:~\footnote{When at least one of the parameters of the superpotential vanishes, we find it impossible to obey the conditions (i) and (ii)
simultaneously.}
\begin{align}
    \mathcal{J}=-3\log{(-C)}-\log(\Phi+\bar{\Phi})~,\\
    \mathcal{W}=\lambda+\mu\Phi+\omega\Phi^2~,
\end{align}
where we parameterize $\Phi$ as
\begin{equation}
    \Phi=y/2+i\theta~,~~~y=e^{-\sqrt{2}\phi}~.
\end{equation}

The scalar potential of the model is given by
\begin{gather}
    V_F=\fracmm{1}{(-C)^3y}\left[\left(-\lambda+\frac{1}{2}\mu y+\frac{3}{4}\omega y^2+\omega\theta^2\right)^2+(\omega y-\mu)^2\theta^2\right]~,\\
    V_D=\fracmm{9g^2}{2}\left(\fracmm{1}{C}+1\right)^2~.
\end{gather}

The critical points can be found analytically as
\begin{gather}
    \theta_0=0~,\\
    y_{0(1)}=-\fracmm{\mu}{9\omega}\left(1\pm\sqrt{1-\fracmm{36}{\mu^2}\lambda\omega}\right)~,~~~y_{0(2)}=-\fracmm{\mu}{9\omega}\left(1\pm\sqrt{1+\fracmm{12}{\mu^2}\lambda\omega}\right)~,\label{yzero}\\
    C_0=-\fracm{1}{2}\left(1+\sqrt{1+\fracm{4B}{A}}\right)~,
\end{gather}
where
\begin{equation}
    A\equiv 3g^2y_0 \quad {\rm and} \quad B\equiv\left(-\lambda+\frac{1}{2}\mu y_0+\frac{3}{4}\omega y_0^2\right)^2~.
\end{equation}

As an example, the signs of the parameters can be fixed as $\omega<0$ and $\lambda,\mu>0$. The points $y_{0(2)}$ lead to a Minkowski vacuum with $V_D=0$, so that in this case $\langle D\rangle=0$  breaks the requirement (i). To exclude $y_{0(2)}$, we can impose the condition $12\lambda|\omega|>\mu^2$ when $y_{0(2)}$ becomes imaginary. Then the remaining minimum at $y_{0(1)}$ is unique with the "plus" branch according to Eq.~\eqref{yzero},
\begin{equation}
    y_0=\fracmm{\mu}{9|\omega|}\left(1+\sqrt{1+\fracmm{36}{\mu^2}\lambda|\omega|}\,\right)~,
\end{equation}
where we have renamed $y_{0(1)}$ to $y_0$.

Unfortunately, the mass of $\phi$ vanishes in the vacuum, similarly to the model studied in Appendix~\ref{WZ type}, so that
and we have to introduce the quartic couplings again.
%%%%%%%%%%%%%%%%%%%%%%%%%%%%%%%%%%%%%%%%%%%%%%%%%%%%%%%%%%%%%%%%%%%%%%%%%%%%%%%%%%%%%%%%%%%%%%%%%%%%%%%%%%%%%%%%%%%%%%%%%%%%%%%%%%%%%%%%%%%%%%%%%%%%%%%%%%%%%%%

\section{The $\Delta$-deformation}\label{delta}

Here we demonstrate that the introduction of a small $\Delta$ in Eq.~$\eqref{deom2}$ does not affect the considerations of Sec.~\ref{newFI}. 

First, let us compute the impact of $\Delta\neq 0$ on the scalar potential during inflation. Equation~$\eqref{sol_deom}$ gets modified as
\begin{align}
J\sim \Delta (C-C_0)-\frac{1}{k}\log \frac{1}{3-\Delta}\left(e^{3k(C-C_0)}-\xi_0e^{k\Delta(C-C_0)}\right)
\end{align}
for $|C|\gg 1$. Uner the assumption $|C\Delta|\ll 1$, while keeping $|C|\gg 1$, it reduces to 
\begin{align}
J_{\infty}\equiv -\fracmm{1}{k}\log \fracmm{-\xi_0}{3}
\end{align}
that is exactly the same as Eq.~$\eqref{Jinfty}$, so that there is no effect of $\Delta$. As regards another relevant function $P(C)$, a straightforward calculation yields 
\begin{align}
P_{\infty}=\fracmm{3-2\xi_0e^{kJ_{\infty}}}{k\xi_0e^{kJ_{\infty}}}-2\Delta \fracmm{e^{-kJ_{\infty}}}{k\xi_0}
\end{align}
for $|C|\gg 1$. Here we have the small correction due to $\Delta$ but it is negligible. Thus the key observation that both $J$ and $P$ become constants 
for $|C|\gg 1$ is unchanged, so that the discussion without $\Delta$ in Sec.~5 can be applied as long as $\Delta$ is small like $|C\Delta|\ll 1$. We conclude that 
the small value of $\Delta$ does not affect inflation.

Next, let us consider the impact of $\Delta$ on the vacuum. Once $\Delta\neq 0$ is included, the original vacuum ($\rho=\rho_0$, $\theta=0$, and $C=-1$) gets shifted. It is difficult to obtain an analytic solution to the vacuum conditions. Therefore, we expand the scalar potential around the original vacuum as follows:
\begin{align}
V=V_0+V_i\Phi^i+\frac{1}{2}V_{ij}\Phi^i\Phi^j+\cdots,\label{expansion}
\end{align}
where $V_i$ and $V_{ij}$ are the first and the second derivatives with respect to all scalar fields $\Phi^i=(C,\rho,\theta)$, respectively, which are evaluated 
at $\rho=\rho_0$, $\theta =0$, and $C=-1$, while $V_0$ is the cosmological constant.

The leading terms of $V_{\rho\rho}$, $V_{\theta\theta}$, and $V_{CC}$ are the same as in Eq.~$\eqref{mass_KKLT_2}$. As regards the remaining terms, we 
find their contributions of the order
\begin{align}
&V_{\rho}=O(\Delta^2),\quad V_{C}=O(\Delta),\quad  V_{C\rho}=O(\Delta),
\end{align}
and 
\begin{align}
V_{\theta}=V_{\theta\rho}=V_{\theta C}=0.
\end{align}

Therefore, the deviations from the original vacuum are 
\begin{align}
\delta \rho\sim O(\Delta^2), \quad \delta \theta=0, \quad \delta C\sim O(\Delta),\label{deviation}
\end{align}
so that they can be safely neglected for small $|\Delta|\ll 1$.

Finally, the cosmological constant $V_0$ is given by
\begin{align}
V_0=-\fracmm{A^2 B^2 e^{3-2 B \rho_0}}{6 \rho_0}+\Delta^2\left(\fracmm{g^2}{16\rho_0}+\fracmm{A^2 B^2 e^{J(-1)-2 B \rho_0}}{18\rho_0(3+\xi_0e^{kJ(-1)})}\right).
\end{align}
Though the obtained correction due to $\Delta$ does uplift the AdS vacuum, it is apparently insufficient to get a dS vacuum because the value of $\Delta$ is supposed to be small as $|\Delta| \ll 1$, while we have the hierarchy of the parameters $|\Delta| \ll A < g$ in our model in accordance to the footnote 6 and 
 Eq.~$\eqref{deviation}$. Moreover, a large value of $\Delta$ to be compatible with $A$ may induce tachyonic masses in $V_{ij}$, see Eq.~$\eqref{expansion}$. Therefore, another mechanism is needed to realize a dS vacuum in our approach, but it is beyond the scope of this paper (see, however, Refs.~\cite{Aldabergenov:2017bjt,Aldabergenov:2019aag} for possible solutions).

\end{appendix}
%%%%%%%%%%%%%%%%%%%%%%%%%%%%%%%%%%%%%%%%%%%%%%%%%%%%%%%%%%%%%%%%%%%%%%%%%%%%%%%%%%%%%%%%%%%%%%%%%%%%%%%%%%%%%%%%%%%%%%%%%%%%%%%%%%%%%%%%%%%%%%%%%%%%%%%%%%%%%%%%%%%%%%%%%%%%%%%%%%%%%%%%%%

 \end{document}